\documentclass[aps,prb,a4paper, amsfonts, amssymb, amsmath, reprint, showkeys, nofootinbib, twoside]{revtex4-2}

\usepackage[english]{babel}
\usepackage[utf8]{inputenc}

\usepackage{amsthm}
\usepackage{mathtools}
\usepackage{physics}
\usepackage{xcolor}
\usepackage{graphicx}
\usepackage{adjustbox}
\usepackage{placeins}
\usepackage[T1]{fontenc}
\usepackage{lipsum}
\usepackage{csquotes}

\begin{document}

\title{ Electron States Emerging at Magnetic Domain Wall of magnetic
  semiconductors with strong Rashba effect.}

\author{I. P. Rusinov} \email[Correspondence email address:
]{rusinovip@gmail.com}
\affiliation{Tomsk State University,
Tomsk, 634050 Russia} \affiliation{St. Petersburg State University,
    199034 St. Petersburg, Russia}

\author{V. N. Men'shov} \affiliation{NRC Kurchatov Institute, Kurchatov
  Sqr. 1, 123182 Moscow, Russia} 

\author{E. V. Chulkov} \affiliation{St. Petersburg State University,
  199034 St. Petersburg, Russia} \affiliation{Donostia International
  Physics Center (DIPC), 20018 San Sebasti\'{a}n/Donostia, Spain}
\affiliation{ Departamento de Pol\'{i}meros y Materiales Avanzados:
  F\'{i}sica, Qu\'{i}mica y Tecnolog\'{i}a, Facultad de Ciencias
  Qu\'{i}micas, Universidad del Pa\'{i}s Vasco UPV/EHU, 20080 San
  Sebasti\'{a}n/Donostia, Basque Country, Spain }

\date{\today} % Leave empty to omit a date

\begin{abstract}
In the present article, we explore the electron properties of magnetic
semiconductors with strong Rashba spin-orbit coupling taking into
account the presence of domain walls at the sample surface. We
consider antiphase domain walls separating domains with both in-plane
and out-of-plane magnetization as well as noncollinear domain
walls. First, we propose the model and unveil general physical picture
of phenomenon supported by analytical arguments. Further, we perform a
comprehensive tight-binding numerical calculations to provide a
profound understanding of our findings. A domain wall separating
domains with any polarization directions is demonstrated to host a
bound state. What is more interesting is that we predict that either
of these domain walls also induces one-dimensional resonant state. The
surface energy spectrum and spin polarization of the states are highly
sensitive to the magnetization orientation in the adjacent
domains. The spectral broadening and the spatial localization of the
resonant state depend significantly on a relation between the Rashba
splitting and the exchange one. Our estimation shows that chiral
conducting channels associated with the long-lived resonant states can
emerge along the magnetic domain walls and can be accessed
experimentally at the surface of BiTeI doped with transition metal
atoms.
\end{abstract}

%\keywords{magnetic topological insulators, band topology}

\maketitle

\section{INTRODUCTION}

In designing next generation of the spintronic technologies, two novel
approaches to non-volatile memory and information processing can be
highlighted. In one of these, electrically manipulated magnetic domain
wall (DW), rather than domains, serves as an active
element~\cite{Parkin_2008,Parkin2015}. Another approach is based on
generation of spin polarized dissipationless
current~\cite{RevModPhys.95.011002}. Recent theoretical
studies~\cite{RevModPhys.95.011002,Varnava2021,Zhao2023} and
experimental findings~\cite{Yasuda2017_MTI,Zhao2023} have predicted
and demonstrated that these approaches can be successfully combined
with each other in both magnetically doped topological insulator (TI)
(Bi,Sb)$_2$Te$_3$ thin films and intrinsic antiferromagnetic TI
MnBi$_2$Te$_4$ flakes, where DWs host robust in-gap bound electron
states.  These topologically protected states may impact on the
realization of peculiar quantum phenomena such as quantum anomalous
Hall effect and axion insulator phase~\cite{RevModPhys.95.011002}.
The linear dispersion relation, spin-momentum locking and quantized
conductivity are special properties of these states. However, their
characteristic features such as group velocity, spin polarization
direction and spatial localization are determined by both the
electrostatic conditions at the TI surface or the interfaces between
TI and trivial insulator and the relative orientation of the
spontaneous magnetization in the domains separated by
DW~\cite{Menshov2023, Menshov2019, PhysRevB.106.205301,
PhysRevB.104.035411, Menshov2021, PhysRevB.99.115301}.  Thus, in
semiconductor materials with topologically nontrivial band structure,
network of the unique channels for chiral electron waves can be
created and manipulated via the surface/interface potential and the
spatially-varying magnetization texture~\cite{Menshov2023,
Menshov2019, PhysRevB.106.205301, PhysRevB.104.035411, Menshov2021,
PhysRevB.99.115301}. In turn, an accumulation of carriers at the bound
state makes possible to move a magnetic DW together with the
conducting channel using an external electric field.

A natural question is whether there are other solids with electron
properties similar to those of magnetic TIs. Among these candidates,
materials containing high-Z (heavy) elements where strong spin–orbit
coupling (SOC) and spatial inversion symmetry breaking are of
prominent interest. According to the Rashba
predictions~\cite{Bychkov_Rashba_1984, Manchon2015, Bihlmayer2022},
owing to the structural inversion asymmetry, electrons propagating
along the surface or interface naturally experience an embedded
potential gradient perpendicular to the boundary. Hereby, even in
nonmagnetic system, the spin-degeneracy of the two-dimensional (2D)
band structure is lifted out, without violating the time reversal
symmetry. This symmetry can be broken by doping with magnetic
impurities or by designing magnetic heterostructures. It bears
intriguing consequences for electrical transport properties of
materials wherein the Rashba splitting of the surface states
intertwines with an exchange splitting. To date there are two
prominent examples of real diluted ferromagnetic (FM) semiconductor
with the highest observed Rashba SOC: Mn-doped
GeTe~\cite{Krempask2016, PhysRevX.8.021067} and V-doped
BiTeI~\cite{Klimovskikh2017, Shikin_2017+bitei+v, Shikin2021}. The
works~\cite{Krempask2016} have experimentally witnessed the
entanglement and manipulation of the magnetic order and SOC splitting
in multiferroic Rashba semiconductor Mn- doped GeTe and the presented
direct evidence for the existence of a strong magnetoelectric coupling
between the electric polarization and the magnetization in this
material. Such a coupling can result in unconventional spin textures
depending on electric biasing under switching, which readily indicates
that the (Ge,Mn)Te samples are far from the ideal monodomain phase.
Transport measurement on (Ge,Mn)Te film has demonstrated that the
Berry curvature dominantly contributes to the anomalous Hall
conductivity~\cite{Yoshimi_2018}. As for V-doped BiTeI, the
angle-resolved photoemission spectroscopy (ARPES)
studies~\cite{Klimovskikh2017, Shikin_2017+bitei+v, Shikin2021} have
demonstrated the giant Rashba-type splitting and the huge exchange gap
of about 90 meV for Bi$_{0.985}$V$_{0.015}$TeI at the
$\overline{\Gamma}$-point. Thereat such a sizable gap does not
correspond to the observed very weak total out-of-plane
magnetization~\cite{Shikin_2017+bitei+v}. That can be explained by the
spontaneous formation of the magnetic domains with opposite spin
polarization below the Curie temperature.

The surface of the as-grown diluted magnetic semiconductors with
strong Rashba SOC can be highly inhomogeneous, with multi-domain
structure~\cite{PhysRevX.8.021067, Shikin_2017+bitei+v}.  It consists
of fragments distinguished by magnitude and orientation of electric
and magnetic polarization as well as spatial scale and shape. These
materials can display electronic properties, which dramatically differ
from those of the ideal uniform counterpart. The main reason for this
is thought to be boundaries between the fragments, i.e., DWs
separating ferroelectric or FM domains. It raises the question of
whether DW can support a spatial confinement of electron density and
how it affects the electronic spin-polarized transport.  However, an
influence of DWs on electron states and the corresponding contribution
to the transport and other properties of the magnetic Rashba systems
remain still poorly understood. In the present article we provide
theoretical study of electron states in the presence of a nonuniform
exchange field caused by a single magnetic DW at the surface of the FM
material with the Rashba SOC. Such a special case has not been solved
so far, seriously hampering the analysis and interpretation of
experimental data. Here, by combining the analytic formalism and
detailed numerical calculations, we thoroughly investigate the
spin-resolved spectral and spatial characteristics of the Rashba
semiconductors with magnetic order. It is found that the DW texture
creates two different types of the quasi-one-dimensional (1D)
spin-polarized electron states in terms of their properties. First, a
purely bound state confined to DW is formed. It has a quasi-parabolic
dependence of energy vs momentum near the 2D surface band continuum.
Second, a peculiar resonant state with a quasi-linear spectrum and a
spin-momentum locking appears as a prominent redistribution of an
electron density around DW. We establish the universality of the
resonant state that emerges in the class of materials under
consideration irrespective of the vector texture of magnetic DW. Along
with that, the velocity and spin polarization of such a state
non-trivially depends on both mutual magnetization orientation in
adjacent domains and orientation relative to the surface. The
DW-induced resonant state and its features resemble a canonical
situation in magnetic TI, where a bound edge state emerges at DW for
the topological reason~\cite{PhysRevB.104.035411, Menshov2021,
Menshov2023, PhysRevB.103.235142}. The dissimilarity proves to be
crucial at relatively large exchange splitting, when the resonant
state energy is strongly broadened due to coupling to the 2D
quasiparticle continuum. In the context of the transport properties,
one should expect that the magnetic DWs play no lesser important role
in the Rashba-type semiconductors than in TIs~\cite{Checkelsky2012}.
Among conceptual consequences, it implies a certain extension of the
principle of bulk-boundary correspondence. Our investigation proposes
a new physical insight on quantum transport in magnetic spin-orbit
materials.

The rest of the paper is organized as follows. In Section II we
introduce the model combining effects of Rashba SOC and a DW-type
magnetization distortion on the electronic properties of a magnetic
semiconductor. By the example of a single antiphase DW we give a
general understanding of the origin of the states at DW and
analytically estimate their spectral features.  In Section III, we
perform comprehensive tight-binding calculations to provide a profound
insight into momentum- and spin-resolved spectral density of both 2D
and 1D states. We investigate the bound and resonant electron states
appearing at sharp interfaces between domains with opposite
out-of-plane (Subsection IIIA) and in-plane (Subsection IIIB)
magnetizations.  Similarly, we address the situation with a
noncollinear DW structure (Subsection IIIC). Section IV contains
detailed discussion on the possibility of experimental observation of
the resonant states in real magnetic semiconductors with strong Rashba
effect in particular BiTeI. We summarize the main results in Section
V.

\section{Model and general relations: Analytical treatment.}

\begin{figure}[t]
\includegraphics[width=\linewidth]{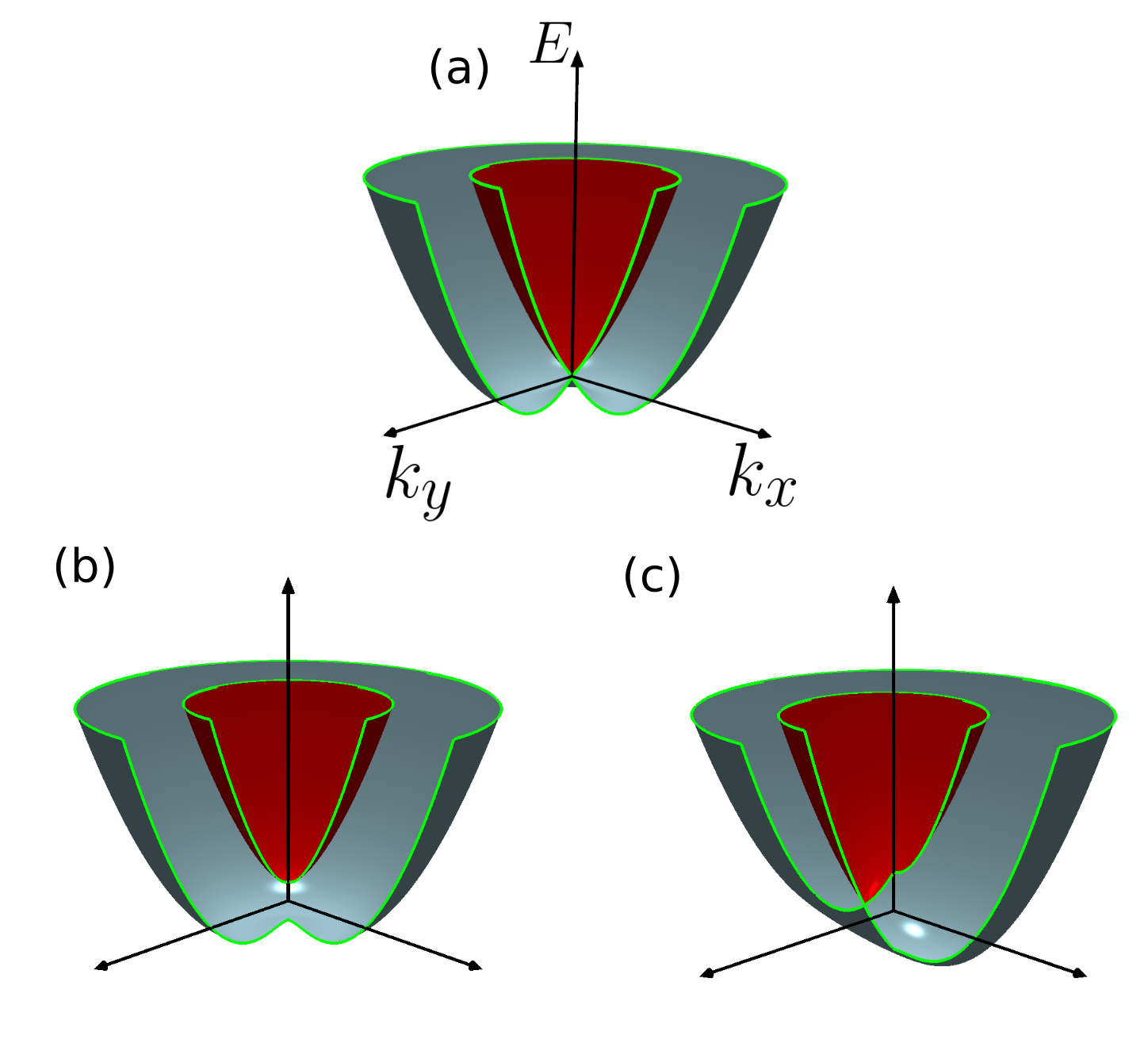}
\caption{
Band structure of the semiconductor surface with spin-orbit Rashba
splitting in three different scenarios: no magnetic order (a),
subjected to the uniform exchange field with out-of-plane direction
 (b) and in-plane direction (c).}
\label{fig:main}
\end{figure}

We study a system of 2D Rashba electrons appearing at the surface of a
three-dimensional semiconductor due to inversion symmetry
violation. These electrons experience an exchange field which may be
caused by either localized moments in the semiconductor host or a
magnetic material layer interfaced with the semiconductor via
proximity effect. Our analysis is based on the minimal model of
electrons moving along the surface $(x,y)$ and affected by both SOC
and exchange field~\cite{Bihlmayer_2015, Manchon2015, Bihlmayer2022},
which is described by the 2D low energy Hamiltonian:
\begin{equation}
H(\mathbf{k}) = \beta \mathbf{k}^2 \sigma_0 -
    \alpha (\left[ \mathbf{k}\times\boldsymbol{\sigma} \right]\cdot 
    \mathbf{e}_z) +
J (\mathbf{M}\cdot\boldsymbol{\sigma}),
\label{eq:main}
\end{equation}
where $\mathbf{k}=(k_x,k_y)$ is the momentum,
$\boldsymbol{\sigma}=(\sigma_x,\sigma_y,\sigma_z)$ is the Pauli matrix
vector, $\sigma_0$ is a nit 2$\times$2 matrix. The first term,
expanded to quadratic order in $\mathbf{k}$ around the
$\overline{\Gamma}$-point (the 2D Brillouin zone center), represents
an electron kinetic energy, where the coefficient $\beta=1/2m^{*}$ is
inversely proportional to the effective mass $m^{*}$. We will set
$\hbar=1$ everywhere. In the vector product term, $\alpha$ is the
Rashba SOC parameter, $\mathbf{e}_z$ is the direction of the SOC field
along which inversion symmetry is broken. The electron spins are
coupled to magnetization $\mathbf{M} = (M_x, M_y, M_z)$ via an
exchange integral $J$, which is assumed to be isotropic and, for
definiteness, positive. The parameter $\alpha$ is uniform in the
$(x,y)$ plane, while the magnetization $\mathbf{M}=\mathbf{M}(x,y)$ is
in general position-dependent. In a locally homogeneous surface
region, where the magnetization $\mathbf{M}$ does not vary, the energy
spectrum of the Hamiltonian~(\ref{eq:main}) is given by
$E_{\pm}(\mathbf{k}) = \beta k^2 \pm \sqrt{J^2 M^2_z + (\alpha k_x -
JM_y)^2 + (\alpha k_y + JM_x)^2}$. Figure~\ref{fig:main} illustrates
the dispersion $E_{\pm}(\mathbf{k})$ for a nonmagnetic case
(Fig.~\ref{fig:main}.a) as well for the out-of-plane
(Fig.~\ref{fig:main}.b) and in-plane (Fig.~\ref{fig:main}.c) the
magnetization orientations. Note that, in experimental situations, the
2D band structure $E_{\pm}(\mathbf{k})$ of the semiconductor surface
can be controlled by electric gating as well as by external magnetic
field applied to the sample.

To learn the role of a magnetic texture in the surface state
modification, let us for simplicity restrict ourselves to a discussion
on Rashba fermions coupled to an exchange filed of a single magnetic
DW approximated with the piece-wise constant profile
$\mathbf{M}(x)=\mathbf{M}^{(r)}h(x)+\mathbf{M}^{(l)}h(-x)$ where
$h(x)$ denotes the Heaviside function. In the following, we use $(r)$
and $(l)$ to label the model parameters for the right and left
semi-infinite domains, respectively. The DW line runs in the $\mathbf{e}_y$
direction for concreteness. The magnetization of both domains is
assumed to remain in $(x,z)$ plane, i.e. $M^{(r)}_y=M^{(l)}_y=0$.
Such magnetization $\mathbf{M}(x)$ breaks translation symmetry along
the $\mathbf{e}_x$ direction and can cause the appearance of peculiar
1D states. The solutions of the corresponding problem,
$H(x,k_y)\Theta(x,k_y) = \epsilon(k_y)\Theta(x,k_y)$, can be
constructed by suitably matching at the interface $x=0$ the spinor
components of the envelope wave-function $\Theta(x,k_y)$ inherent to
homogeneous domain regions around DW. The eigen energies and functions
of the problem depend parametrically on the momentum $k_y$ that is
preserved. For each value of $\epsilon$ there are four characteristic
wave-vectors which are given by
\begin{widetext}
\begin{equation}
\begin{aligned}
[q^{(r)}_{1,2}(k_y,\epsilon)]^2&  =  k^2_y - \frac{1}{2\beta^2}
\left[
2\beta\epsilon+\alpha^2 \mp 
\sqrt{\alpha^4 + 4\alpha^2\beta\epsilon +
4\beta^2(\Delta^{(r)2}_z+\Delta^{(r)2}_x+2\alpha k_y\Delta^{(r)}_x)}
\right], \\
[q^{(l)}_{1,2}(k_y,\epsilon)]^2&  =  k^2_y - \frac{1}{2\beta^2}
\left[
2\beta\epsilon+\alpha^2 \mp 
\sqrt{\alpha^4 + 4\alpha^2\beta\epsilon +
4\beta^2(\Delta^{(l)2}_z+\Delta^{(l)2}_x-2\alpha k_y\Delta^{(l)}_x)}
\right],
\end{aligned}
\end{equation}
\end{widetext}
where $\Delta^{(r,l)}_{x,z} = JM^{(r,l)}_{x,z}$. For the sake of
definiteness, we assume $\Delta^{(r,l)}_{x,y} > 0$. The linear
combination of the corresponding exponential functions, $\sim\exp[\pm
q^{(r,l)}_{1,2}(k_y,\epsilon)x]$, which satisfies the boundary
conditions, provides the sought solution.

To have reference points, we begin with the study of the case of an
easy-axis anisotropy, when the out-of-plane magnetization
$\mathbf{M}=\mathbf{e}_z M_z$ ($M_x=M_y=0$) is preferred in the ground
state. The system can be characterized by two energy scales, namely,
the Rashba SOC splitting,
$E_{\text{so}}=k^2_{\text{so}}/2m^{*}=\alpha^2/4\beta$, being the
energy difference between the crossing point and the band minimum at
$\mathbf{M}=0$, $k_{\text{so}}=m^{*}\alpha=\alpha/2\beta$ is the
minimum position in the momentum space. The magnetization
$\mathbf{M}=\mathbf{e}_z M_z$ lifts the spin degeneracy at the Kramers
point and opens the exchange gap $2\Delta_z=2JM_z$ at $\mathbf{k}=0$
between the upper and lower spin-split subbands, $E_{\pm}(k)=\beta
k^2 \pm \sqrt{\Delta_z^2+\alpha^2 k^2}$, in the spectrum of the
Hamiltonian~(\ref{eq:main}) (see Fig.~\ref{fig:main}.b). The lower
subband exhibits one minimum or two minima depending on whether the
system is in the exchange-dominated regime, $\Delta_z >
2E_{\text{so}}$, or in the SOC-dominated regime, $\Delta_z <
2E_{\text{so}}$. When $\Delta_z > 2E_{\text{so}}$ and $\epsilon <
-\Delta_z$, $[q_1^{(r,l)}(k_y,\epsilon)]^2$ and
$[q_2^{(r,l)}(k_y,\epsilon)]^2$ are positive. In turn, when
$2E_{\text{so}} > \Delta_z$ and $\epsilon <
-E_{\text{so}}(1+(\Delta_z/2E_{\text{so}})^2)$,
$[q_1^{(r,l)}(k_y,\epsilon)]^2$ and $[q_2^{(r,l)}(k_y,\epsilon)]^2$
are complex conjugated each other. In both cases, the bound electron
state (indexed with 'B') with the energy
$\epsilon=\epsilon^{(\perp)}_B(k_y)$ below the conduction band bottom
occurs at the out-of-plane DW ['$(\perp)$'] on the localization scale
$\sim|\text{Re}\ q^{(r,l)}_{1,2}(k_y,\epsilon)|^{-1}$. In particular,
in the case of the antiphase DW with profile
$\mathbf{M}(x,y)=\mathbf{e}_z M_z \text{sgn}(x)$, when
$q^{(r)}_{1,2}(k_y,\epsilon) =
q^{(l)}_{1,2}(k_y,\epsilon)=q_{1,2}(k_y,\epsilon)$, there appears the
state with the dispersion relation $\epsilon^{(\perp)}_B(k_y)$, which
is asymmetric with respect to $k_y=0$~\cite{PhysRevResearch.2.022052}.
At $k_y=0$ the energy $\epsilon^{(\perp)}_B(0)$ satisfies the equation
$\beta\epsilon\Big\{[q_1(0,\epsilon)]+[q_2(0,\epsilon)]\Big\}^2+\Delta^2_z=0$.
Interestingly, when $\Delta_z/2E_\text{so} \ll 1$, the dispersion
curve $\epsilon^{(\perp)}_{B}(k_y)$ approaches the energy level
$-E_{\text{so}}$ occupying the interval $|k_y| < k_{\text{so}}$.  In
this limit, one obtains
$\epsilon^{(\perp)}_B(0)=-(E_{\text{so}}+\Delta^2_z/2E_\text{so})$ and
$q_{1,2}(0,\epsilon^{(\perp)}_B(0))=k^{(\perp)}\pm i k_\text{so}$
(where $k^{(\perp)}=\Delta_z/\alpha$), in other words, the envelope
function decays slowly away from DW and simultaneously quickly
oscillates. Thus, the magnetic DW acts as a short-range linear defect
for Rashba’s fermions, which causes the bound state
$\epsilon^{(\perp)}_B(k_y)$ below the 2D band continuum.

Apart from the bound electron state, we are interested in another
specific type of 1D electron state, which can arise at the
out-of-plane magnetic DW. The energy $\epsilon$ of this state lies
inside the local exchange gap. For definiteness, we will talk about
the antiphase DW, $\mathbf{M}(x,y)=\mathbf{e}_zM_z\text{sgn}(x)$.
Note that, in the energy range $E_{+}(k_y)>\epsilon>E_{-}(k_y)$, the
value $q_1(k_y,\epsilon)$ is real, while the value $q_2(k_y,\epsilon)$
is imaginary. Therefore, on the one hand, the spinor wave with the
wave-vector $q_2(k_y,\epsilon)$ propagates along the surface and,
having reached the magnetic DW, is partly transmitted through the
interface or partly reflected from it. On the other hand, on each side
of the DW, there is an evanescent wave $\sim\exp[\pm
q_1(k_y,\epsilon)x]$. The full envelope function is expressed as a
superposition of the above propagating and evanescent waves. It
describes an electron state that is placed within the local exchange
gap and degenerate with the continuum of the lower subband
$E_{-}(k_y)$. So one can consider this state as a resonant one ('R').
Within the scattering theory, the characteristics of the resonant
state are contained in the reflection and transmission coefficients,
which can be analytically estimated for the sharp DW. Under conditions
of a relatively weak exchange energy, $\Delta_z \ll 4E_\text{so}$,
cumbersome calculations give us the expression for the reflection
coefficient as a function of energy and momentum
\begin{equation}
R^{(\perp)}(\epsilon,k_y)=\frac{\Delta_z}{4E_\text{so}}
\frac{\epsilon+\text{Re}[\epsilon^{(\perp)}_R(k_y)]}
{\epsilon-\epsilon^{(\perp)}_R(k_y)},
\label{eq:refl}
\end{equation}
where $\text{Re}[...]$ denotes the real part.

Writing down Eq.~(\ref{eq:refl}) under the stipulation
$\Delta_z/4E_\text{so} \ll 1$ and $|k_y|/k_\text{so} \ll 1$ we keep
main non-vanishing terms in the dispersion relation
\begin{equation}
\epsilon^{(\perp)}_R(k_y) = \alpha k_y + \omega^{(\perp)}_R - i \Gamma^{(\perp)}_R.
\label{eq:4}
\end{equation}
In this limit, the resonant state inside the local exchange gap,
$|\epsilon^{(\perp)}_R(k_y)|<\Delta_z$, is specified by linear
momentum-energy dependence with the group velocity $\alpha$, small
energy shift upwards $\omega^{(\perp)}_R = \Delta^2_z/4E_{\text{so}}$
and scanty broadening width $\Gamma^{(\perp)}_R
= \Delta^3_z/8E^2_\text{so}$. Such a smallness of $\Gamma^{(\perp)}_R$
justifies the expansion (4) for the resonance. A notable feature is
that the resonance narrows significantly faster than the exchange gap
closes. Near the resonance, the squared reflection probability takes
the Lorentzian form,
$|R^{(\perp)}(\epsilon,0)|^2 \approx \frac{(\Gamma^{(\perp)}_R)^2}
{(\epsilon-\omega_R^{(\perp)})^2+(\Gamma^{(\perp)}_R)^2}$, with the
distance from the resonance, it goes to the value
$|R^{(\perp)}(\epsilon,0)|^2 \approx (\Delta_z/4E_{\text{so}})^2$, and
it passes through zero (an antiresonace) at
$\epsilon=-\omega^{(\perp)}_R$. The spatial distribution of the
envelope function of the state (4) exhibits the significant localized
component around DW, $\sim a_1 \exp(\pm k^{(\perp)}|x|)$, and the
propagating wave of a minor magnitude, $\sim a_2\exp(\pm 2ik_\text{so}
x)$, i.e. $|a_2/a_1|^2 \sim \Delta_z/E_\text{so}$. Thus, the in-gap
resonant state acquires a quasi-bound character in the direction
perpendicular to DW and propagates along DW with the slight
attenuation. Moreover, the state is highly spin-polarized along the
$\mathbf{e}_x$ axis.

Similarly to the above analysis one can address the case of an
easy-plane anisotropy. The in-plane magnetization does not open an
energy gap in the Rashba spectrum, however, it displaces the crossing
point position in momentum space from the Brillouin zone center, as
seen in Fig.~\ref{eq:main}.c. Therefore, for the sharp antiphase DW
with the domain polarization directed normal to DW,
$\mathbf{M}(x,y)=\mathbf{e}_x M_x \text{sgn}(x)$due to the mirror
symmetry with respect to the $(y,z)$plane, the projected 2D bands are
given by the four gapless branches, $E^{(r)}_{\pm}(k_x=0,k_y) = \beta
k^2_y \pm (\alpha k_y + \Delta_x)$ and $E^{(l)}_{\pm}(k_x=0,k_y)=\beta
k_y^2 \pm (\alpha k_y-\Delta_x)$ (where $\Delta_x=JM_x$) which are
derived from each domain in pairs. Note that these dependencies attain
the minimum values at $|k|=k_\text{so}$, namely,
$E^{(r)}_{\pm}(k_x=0,k_y=\pm k_\text{so})=-E_\text{so}\pm \Delta_x$
and $E^{(l)}_{\pm}(k_x=0, k_y =\pm k_\text{so}) =
-E_\text{so} \mp \Delta_x$.  The characteristic feature of the
spectrum projection is an energy-momentum window restricted by the
four branches $E^{(r)}_{\pm}(k_x=0,k_y)$ and
$E^{(l)}_{\pm}(k_x=0,k_y)$, which in the limit
$\Delta_x/4E_\text{so} \ll 1$ acquires the shape of an almost perfect
rhombus. Solving the problem of a low-energy scattering of the Rashba
electrons on the in-plane DW ['$(\parallel)$'], we find that, within the
rhombus-like window, there is a 1D resonant state. Under the condition
$\Delta_x/4E_\text{so} \ll 1$, the dispersion relation for this state
reads
\begin{equation}
\epsilon_{R}^{(\parallel)}(k_y)=
\lambda [k_y^2-(k^{(\parallel)})^2+
\omega^{(\parallel)}_R-i\Gamma^{(\parallel)}_R],
\label{eq:5}
\end{equation}
where $\omega^{(\parallel)}_R=\Delta_x^2/4E_\text{so}$,
$\Gamma^{(\parallel)}_R = \Delta_x^3/8E^2_\text{so}$,
$k^{(\parallel)}=\Delta_x/\alpha$ and $\lambda \approx 4\beta
(\Delta_x/E_\text{so})^2$.  Remarkably, the energy band
$\epsilon^{(\parallel)}_R(k_y)$ exists in a small portion of the 1D
Brillouin zone, $|k_y|<k^{(||)}$, and has an effective mass $m^{*}_R$
which greatly exceeds that of the 2D surface band, $m^{*}_R \approx
(E_\text{so}/\Delta_x)^2 m^{*}$. The width of such a flattened band is
extremely small, $\sim \Delta^4_x/E^3_\text{so}$, i.e., it is narrower
than the resonance broadening $\Gamma^{(\parallel)}_R$ Thus, the heavy
fermion emerging at the DW has the finite lifetime, in addition to the
weak dispersion. Therefore, the state~(\ref{eq:5}) will form a sharp
peak in DOS near zero energy. The resonant state is highly
spin-polarized along the $\mathbf{e}_z$ direction, in other words, it
is almost chiral. And the polarization sign depends on whether the
in-plane DW has a head-to-head texture or a tail-to-tail one.

The in-plane DW also hosts the bound state. One can show that the
antiphase DW, $\mathbf{M}(x,y)=\mathbf{e}_x M_x \text{sgn}(x)$,
creates the state exponentially localized at the interface. It has a
parabolic-like spectrum with a negative effective mass,
$\epsilon^{(\parallel)}_B(k_y)$, which is symmetric with respect to
$k_y=0$ and fixed within the interval $-E_\text{so}-\Delta_x
< \epsilon^{(\parallel)}_B(k_y) <
2\Delta_x-E_\text{so}-\Delta^2_x/4E_\text{so}$.

The essential physics of the origin of the resonant states captured by
our low-energy effective model is the following. Based on the standard
Hamiltonian for magnetic semiconductor with strong Rashba-like SOC, we
modify the exchange part in Eq.~(\ref{eq:main}) by introducing a
spatially varying magnetization, which gives a natural way to segment
the surface into regions with distinct topological indicator
values. Formally, when $\beta \to 0$ or $m^{*}\to\infty$, the
Hamiltonian~(\ref{eq:main}) is of the same form that describes pure
Dirac fermions for the 2D TI, where the constant $\alpha$ represents
their velocity~\cite{hasan2010colloquium}. In such a case, the 1D
bound state with a linear spectrum spanning the exchange gap is known
to be induced by the interface between two magnetic domains with
opposite out-of-plane magnetization directions~\cite{Checkelsky2012}.
In turn, chiral symmetry of the linearized Hamiltonian~\ref{eq:main}
allows for the existence of a state with a flat band at zero energy at
the in-plane
DW~\cite{RevModPhys.88.035005}. Recently~\cite{PhysRevB.104.035411,
Menshov2021}, we have shown that a single antiphase DW on the surface
of the magnetic TI with an easy-plane anisotropy hosts the chiral
dispersionless state with a zero energy. Magnetic DWs with more
complex textures also can create the bound states in
TI~\cite{PhysRevB.104.035411, Menshov2021, PhysRevB.103.235142}.  In
any case, such states (placed inside a global energy gap) are
protected by the change in topological invariant across the magnetic
DW. The similar effect can occur at the magnetic DW in the Rashba
semiconductor. Indeed, under the condition
$\Delta_{x,z}/4E_\text{so} \ll 1$, the dominant contribution to the
formation of the 1D resonant states at DW comes from the 2D Rashba
states with small momenta near $\mathbf{k}=0$, $|k_y| \ll
k_\text{so}$, and our results reproduce those of the TI
theory. However, the role of higher order terms in the momentum grows
with an increase in the exchange field. Taking in account the
quadratic term, $\mathbf{k}^2$, in Eq.~(\ref{eq:main}), the bound
state defined in linearized approximation is transformed into the
resonant (quasi-bound) state with the spectra~(\ref{eq:4}) and
(\ref{eq:5}), which acquires the small energy shift and broadening. In
turn, the envelope function of the resonant state becomes rather
strongly localized near DW within pre-asymptotic region. Under the
weak out-of-plane exchange field, the existence of the in-gap
linear-in-momentum state~(\ref{eq:4}) is protected owing to almost
quantized change of a topological invariant at the interface,
$C^{(r)}-C^{(l)}\approx1-\Delta_z/4E_\text{so}$. The state
(\ref{eq:5}) with flattened spectrum can be attributed to the
approximate chiral symmetry of the model with the in-plane magnetic
anisotropy. The effect of resonance broadening of the 1D in-gap state
$\epsilon^{(\perp,\parallel)}_R(k_y)$ is due to its coexistence with a
continuous branch $E_{-}(\mathbf{k})$ of the 2D Rashba states. As the
exchange splitting increases, the energy shift
$\epsilon^{(\perp,\parallel)}_R$ increases and the resonance becomes
broader, which implies that the state completely loses its meaning in
the limit $\Delta_{x,z}/4E_\text{so} \gg 1$.

\section{Tight-binding simulations. Results.}

To gain our insight into the electron states hosted at the magnetic
DWs in Rashba semiconductor surface, we resort to numerical
simulations based on a tight-binding approximation. This method allows
us to reproduce in detail the features of these states for an
arbitrary spatial profile of the DW texture in the whole region of the
model parameters, which cannot be performed analytically and which is
highly time-consuming within {\it ab initio} technique.

\begin{figure*}[t]
\includegraphics[width=0.9\linewidth]{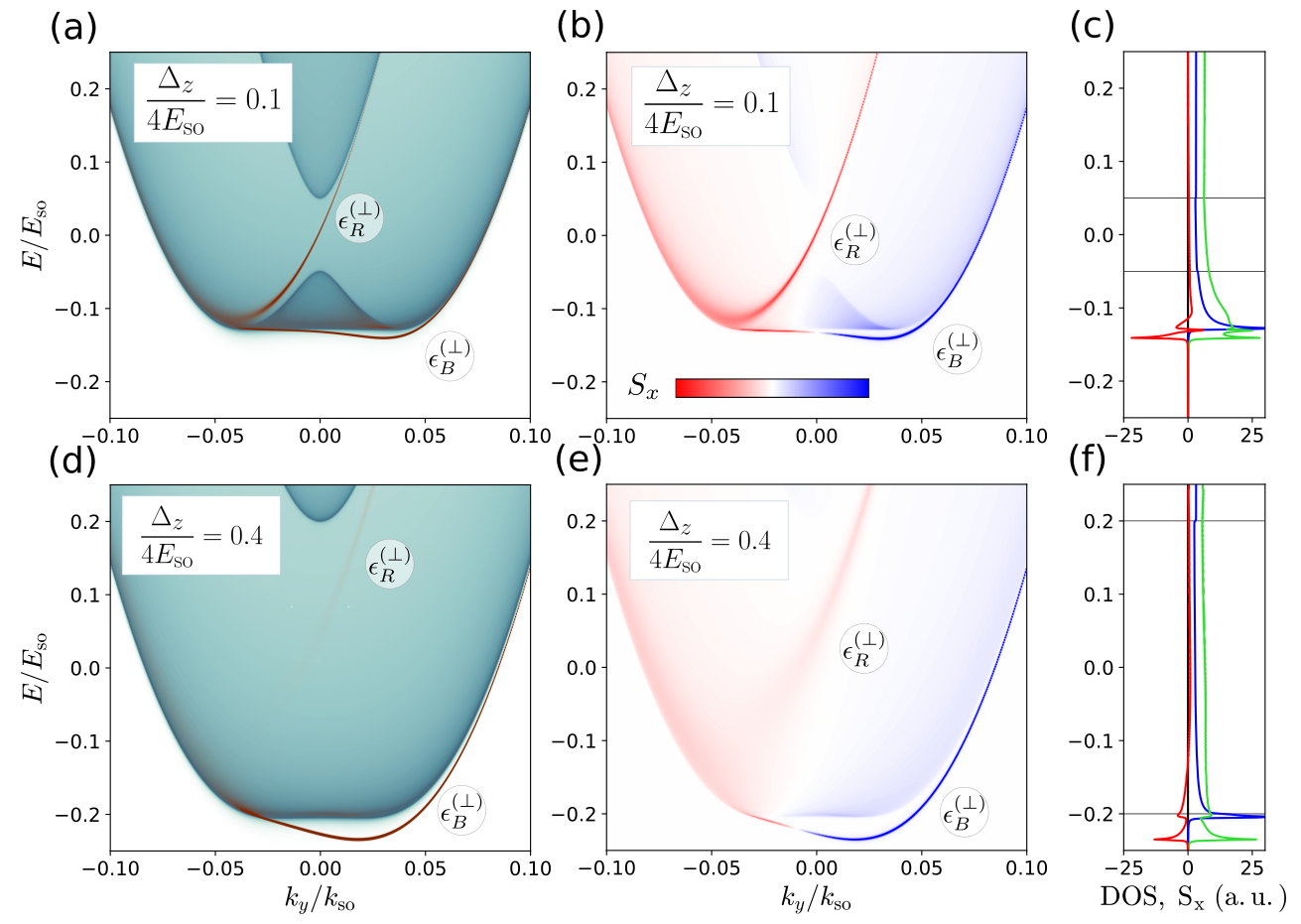}
\caption{
Spectral characteristics of the electron states at the surface of the
magnetic Rashba semiconductor divided into two out-of-plane
domains. (a) and (d) Spectral function intensity. The projection of
the 2D domain states is shown in azure-green color and the 1D states
stemmed from DW are in cherry color. (b) and (e) Spin polarized
spectral function intensity. The positive/negative spin polarization
along the $\mathbf{e}_x$ axis is represented in blue/red color. (c)
and (f) Energy dependence of 2D DOS (blue curve), 1D DOS (green curve)
and the $S_x(E)$ component of spin density (red curve). The bound
state $\epsilon^{(\perp)}_{B}(k_y)$ and resonant state
$\epsilon^{(\perp)}_R$ are indexed in the figure. The upper panels is
for $\Delta_z/4E_\text{so}=0.1$, the lower panels is for
$\Delta_z/4E_\text{so}=0.4$.}
\label{fig:fig2}
\end{figure*}

\begin{figure}[t]
\includegraphics[width=\linewidth]{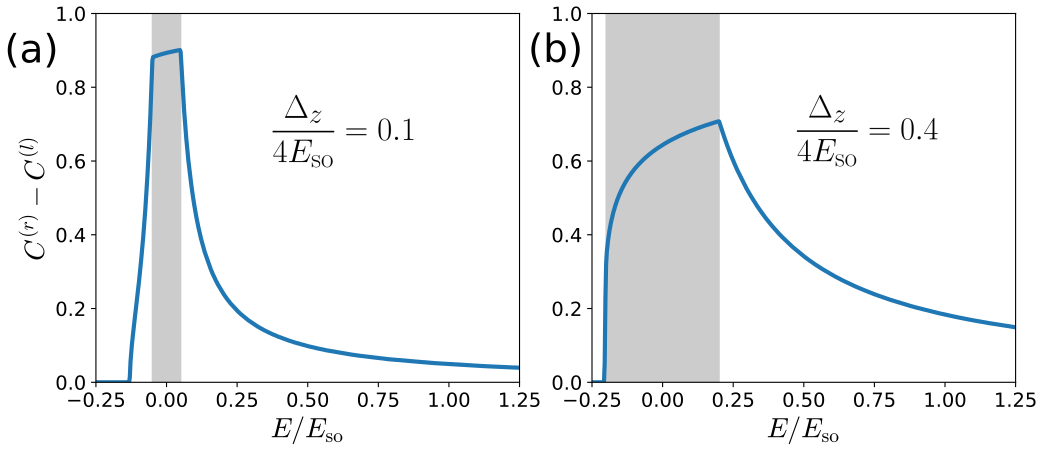}
\caption{
Topological probe of the magnetic Rashba semiconductor surface. (a)
and (b) The difference between the Chern numbers of the right and left
domains $C^{(r)}-C^{(l)}$ as a function of the chemical potential
position $\mu$ for two values of the ratio $\Delta_z/4E_{\text{so}}$,
0.1 and 0.4, respectively.
}
\label{fig:fig3}
\end{figure}

According to tight-binding formalism we perform a lattice
regularization of the continuum Hamiltonian~(\ref{eq:main}) via the
substitution $k_{x,y} \to \frac{1}{a} \sin(k_{x,y}a)$ and
$k^2_{x,y} \to \frac{2}{a^2} [1-\cos(k_{x,y}a)]$, where $a$ is a 2D
square lattice constant and use recursive technique for the Green
functions~\cite{Sancho:85, Henk:93}. Having obtained the retarded
Green function $G(x,x',k_y,E+i\eta)$ for a certain magnetization
distribution, we can calculate the corresponding one-particle spectral
function $A(k_y,E)=-\frac{1}{\pi}\lim_{\eta \to 0+} \Im \Tr \int dx
G(x,x,k_y,E+i\eta)$ and spin polarized spectral function
$S_i(k_y,E)=-\frac{1}{\pi}\lim_{\eta \to 0+} \Im \Tr \int
dx \boldsymbol{\sigma}_i G(x,x,k_y,E+i\eta)/A(k_y,E)$.  The total DOS
and the spin density components in dependence on the state energy $E$
can be calculated from the expressions $\rho(E)
= \frac{2\pi}{a} \int_{-\pi/a}^{\pi/a} dk_y A(k_y,E)$ and $S_i(E)
= \frac{2\pi}{a} \int_{-\pi/a}^{\pi/a} dk_y S_i(k_y,E)$, respectively.
We assume periodic boundary conditions in the $\mathbf{e}_y$
direction, along which the DW is running.

\subsection{Out-of-plane DW structure}\label{sub:oop_dw}

The results obtained from analysis of the spectral functions
$A(k_y,E)$ and $S(k_y,E)$ in the case of the out-of-plane magnetic DW
are demonstrated in Fig.~\ref{fig:fig2}. As has been discussed above,
in the magnetic Rashba semiconductor with perpendicular-to-surface
uniaxial anisotropy the surface electron spectrum exhibits an opening
of the exchange gap $2\Delta_z$, which is clearly reflected in
Fig.~\ref{fig:fig2}.a,b as well as in Fig.~S1 of Supplemental
Materialfor different values of $\Delta_z/4E_\text{so}$.
Figure~\ref{fig:fig2}.c,d exemplifies spin textures of opposite
chirality for the projection of the inner and outer branches of the 2D
spectrum, $E_{+}(k_y)$ and $E_{-}(k_y)$, respectively. In addition, in
Fig.~\ref{fig:fig2}.e,f one can see the van Hove-like singularity in
the vicinity of the conduction band bottom,
$\rho^{(\perp)}_{\text{2D}}(E)
= \frac{a^2}{2\pi\beta}(1+\frac{\Delta_z^2}{4E^2_\text{so}}+\frac{E}{E_\text{so}})^{-1/2}$,
which is typical for DOS of the Rashba system. As can be observed in
Fig.~\ref{fig:fig2}, regardless of the relation between the exchange
splitting and the Rashba one, a separation of the surface into two
domains with opposite magnetization directions,
$\mathbf{M}(x,y)=\mathbf{e}_z M_z \text{sgn}(x)$, leads to an
appearance of the 1D bound state at DW. The energy branch
$\epsilon^{(\parallel)}_{B}(k_y)$ defining this state, is split off
from the band continuum $E_{-}(k_y)$, it is well-defined and clearly
distinguished in the spectral picture of Fig.~\ref{fig:fig2}. Its
dispersion is not symmetric with respect to $k_y=0$, i.e., the largest
energy separation is placed at a finite value of momentum $k_y$. With
an increase in the ratio $\Delta_z/4E_\text{so}$, the minimum of the
dependence $\epsilon^{(\perp)}_B(k_y)$ becomes deeper and moves closer
to the middle of the Brillouin zone (see Fig.~S1). The spin texture of
the bound state, shown in Fig.~\ref{fig:fig2}.c,d under the relatively
small exchange field, has a complicated composition in the momentum
space: the positive (negative) in-plane spin polarization,
$S^{(\perp)}_{xB}(k_y,\epsilon)>0$
($S^{(\perp)}_{xB}(k_y,\epsilon)<0$), occurs mainly near
$k_y\approx-\sqrt{k^2_\text{so}-(k^{(\perp)})^2}$
($k_y\approx\sqrt{k^2_\text{so}-(k^{(\perp)})^2}$). This fact is
manifested as a presence of two peaks with opposite sign in the spin
density $S^{(\perp)}_{xB}(\epsilon)$, which contribute to DOS of the
bound state, $\rho^{(\perp)}_B(\epsilon)$, below the 2D band edge [see
Fig.~\ref{fig:fig2}.e,f].

The most remarkable feature in the weak exchange field regime,
$\Delta_z/4E_\text{so} \ll 1$, plotted in Fig.~\ref{fig:fig2}.a,c,
is the pronounced spectral branch connecting the 2D inner and outer
bands through the gap. It shows the almost linear energy-momentum
dependence within the local exchange gap and hardly visible
broadening. Such a spectral feature is naturally associated with the
in-gap resonant state induced by the out-of-plane magnetic DW, which
is analytically described by Eq.~(\ref{eq:4}) for the dispersion
$\epsilon^{(\perp)}_R(k_y)$. However, with the increasing exchange gap
size, this branch broadens substantially until its signature
disappears at $\Delta_z \gtrsim E_\text{so}$ [see
Fig.~\ref{fig:fig2}.b,d and Fig.~S1; note that spin resolution
allows us to better see the resonance even at
$\Delta_z/4E_\text{so}=0.4$]. Another important feature: the
unidirectional resonant state $\epsilon^{(\perp)}_R(k_y)$ possesses a
nearly perfect spin polarization $S^{(\perp)}_{xR}(k_y,\epsilon)>0$
[Fig.~\ref{fig:fig2}.c-f].  In other words, it is a right-handed
chiral state. Apparently, the out-of-plane DW with a negative
magnetization gradient hosts a left-handed chiral state, because a
change $\mathbf{M}(x,y)=\mathbf{e}_z M_z \text{sgn}(x)$ $\to$
$\mathbf{M}(x,y)=-\mathbf{e}_z M_z \text{sgn}(x)$ leads to inversion
of sign of both the electron velocity
$d\epsilon^{(\perp)}_R(k_y)/dk_y$ and spin polarization
$S^{(\perp)}_{xR}(k_y,\epsilon)$.

\begin{figure*}[t]
\includegraphics[width=0.8\linewidth]{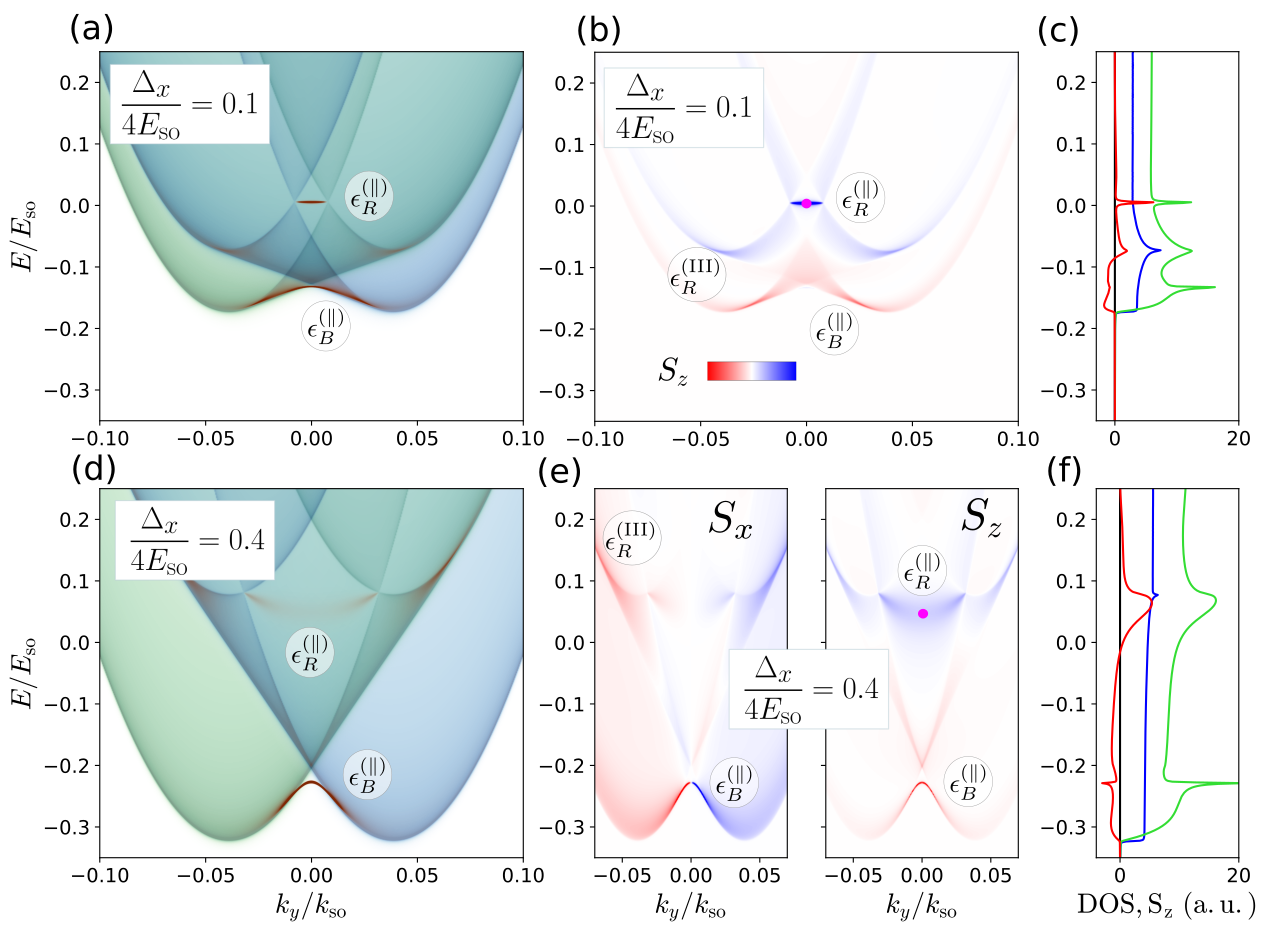}
\caption{
Spectral characteristics of the electron states on the surface of the
magnetic Rashba semiconductor divided into two in-plane domains. (a)
and (d) Spectral function intensity. The projection of the 2D domain
states is shown in azure-green color and the 1D states stemmed from DW
are in cherry color. (b) and (e) Spin polarized spectral function
intensity. The positive/negative spin polarization along the
$\mathbf{e}_x$ axis is represented in blue/red color. (c) and (f)
Energy dependence of 2D DOS (blue curve), 1D DOS (green curve) and
$S_z(E)$ component of spin density (red curve). The bound state
$\epsilon^{(\parallel)}_B(k_y)$ and resonant states
$\epsilon^{(\parallel)}_R(k_y)$ and $\epsilon^{(\text{III})}_R(k_y)$
are indexed in the figure. The upper panels is for
$\Delta_x/4E_\text{so}=0.1$, the lower panels is for
$\Delta_x/4E_\text{so}=0.4$.  }
\label{fig:fig4}
\end{figure*}

\begin{figure}[t]
\includegraphics[width=0.8\linewidth]{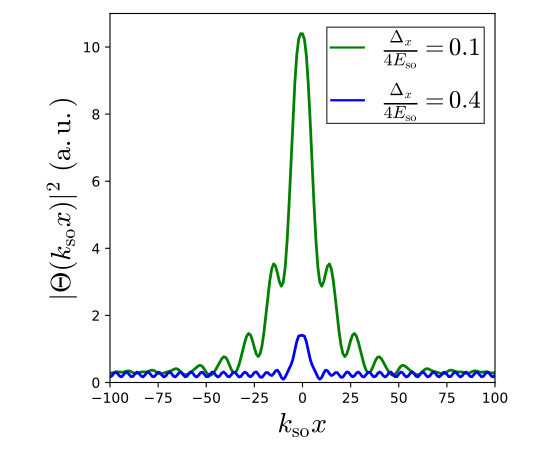}
\caption{
Probability density for the resonant state
$\epsilon^{(\parallel)}_R(k_y)$, which corresponds with the red violet
points in Fig.~\ref{fig:fig4}.b,e. The ratio $\Delta_x/4E_\text{so}$
has been taken as 0.1 in green and 0.4 in blue.}
\label{fig:fig5}
\end{figure}

\begin{figure*}[t]
\includegraphics[width=0.9\linewidth]{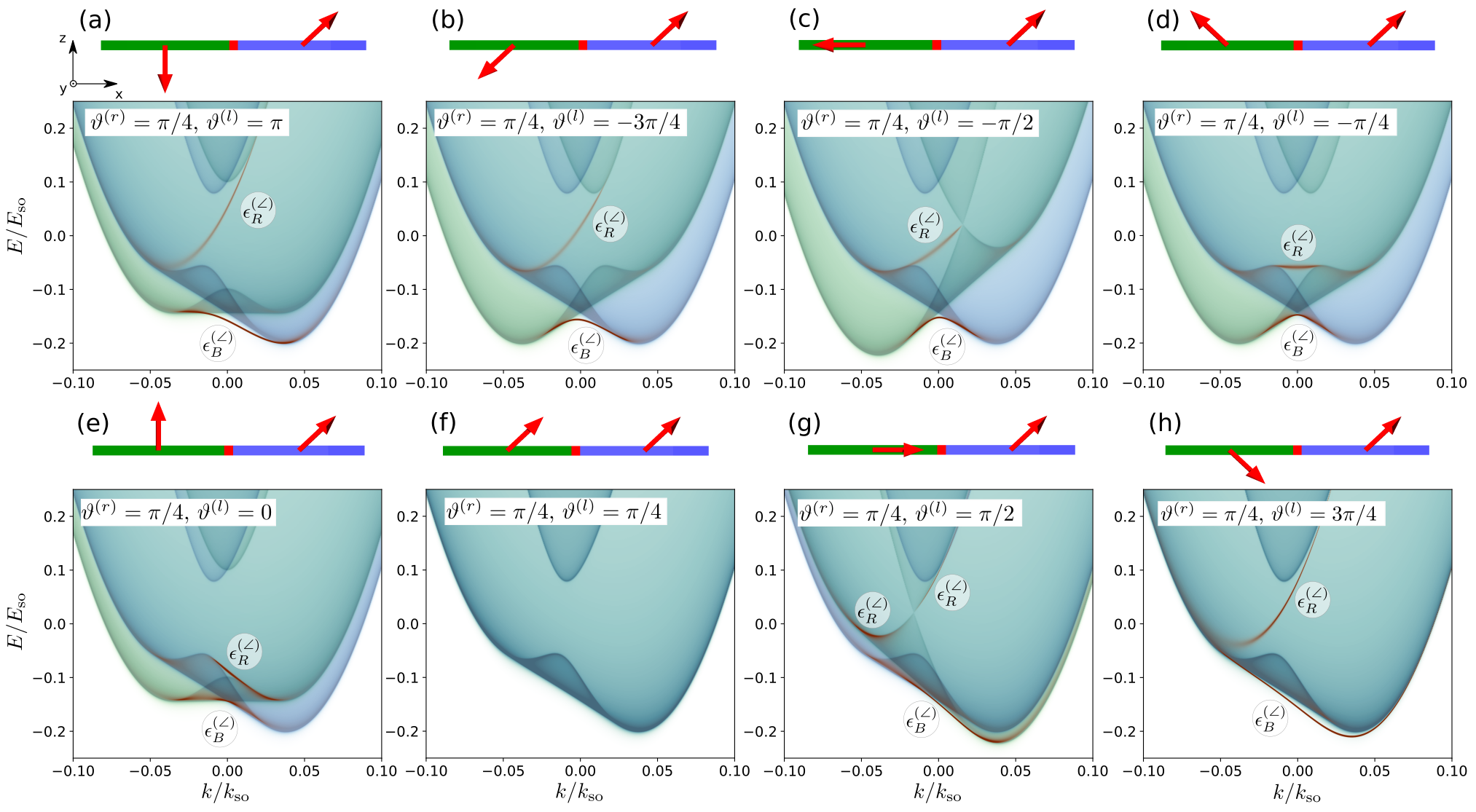}
\caption{
Spectral function intensity of the electron states at the surface of
the magnetic Rashba semiconductor divided into two domains with
noncollinear magnetizations $\mathbf{M}^{(r)}$ and $\mathbf{M}^{(l)}$
confined in the $(x,z)$ plane. It is possible to follow the evolution
of the energy spectrum when the vector $\mathbf{M}^{(l)}$ rotates
while the direction of $\mathbf{M}^{(r)}$ is fixed
($\vartheta^{(r)}=\pi/4$). The projection of the 2D domain states is
made in azure-green color and the 1D states stem from DW are in cherry
color. The bound state $\epsilon^{(\angle)}_B(k_y)$ and the resonant
states $\epsilon^{(\angle)}_R(k_y)$ are indexed in the figure. In the
upper part of the panels, the red arrows indicate the magnetization
directions in the right (blue ribbon) and left (green ribbon)
domains. The ratio $\Delta_0/4E_\text{so}$ used in these plots is 0.2.
}
\label{fig:fig6}
\end{figure*}

The behavior of the resonant state under variation in the parameters
of the Hamiltonian~(\ref{eq:main}) may be related to its topological
properties.  In the case of the antiphase out-of-plane DW, it is
natural to consider the difference between the Chern numbers of the
right and left domains, $C^{(r)}$ and $C^{(l)}$, respectively. The
dependence $C^{(r)}-C^{(l)}$ as a function of the chemical potential
position $\mu$ is shown in Fig.~\ref{fig:fig3}. One can see that
this dependence changes markedly with variation in the ratio
$\Delta_z/4E_\text{so}$.  When the chemical potential is fixed inside
the exchange gap, $|\mu|<\Delta_z$, one obtains:
\begin{equation}
C^{(r)}-C^{(l)}=1-\frac{\Delta_z}{\sqrt{\Delta_z^2+\Omega^2(\mu)}},
\label{eq:6}
\end{equation}
\begin{equation}
\Omega^2(\mu) = 8E^2_\text{so}\Big[
1+\frac{\mu}{2E_\text{so}}+
\sqrt{
\Big(1+\frac{\mu}{2E_\text{so}}\Big)^2+
\frac{\Delta_z^2-\mu^2}{4E^2_\text{so}}
}
\Big]
\end{equation}
The adjusted domains have contrasting topological properties in the
reciprocal space which, in the weak exchange field regime,
$\Delta_z/4E_\text{so} \ll 1$, results in what the Chern number
undergoes the change closer to an unit quantity. This change
associated with the out-of-plane magnetic DW gives rise to the
appearance of the in-local-gap resonant state. However,
at large exchange field, namely, $\Delta_z/4E_\text{so} \gtrsim 1$,
the difference $C^{(r)}-C^{(l)}$~(Eq.~(\ref{eq:6})) becomes
poorly quantized, as a consequence, the resonance is being eroded
because its width significantly increases.

\subsection{In-plane DW structure}

The spectral function of the Rashba fermions coupled to the in-plane
magnetic DW of a tail-to-tail type,
$\mathbf{M}(x,y)=\mathbf{e}_xM_x\text{sgn}(x)$, contributed from 2D
and 1D surface electron states is presented in
Fig.~\ref{fig:fig4}. Figure~S2, in addition to
Fig.~\ref{fig:fig4}, allows us to imagine the evolution of the
system band structure with increasing the exchange field size
$\Delta_x=JM_x$. The 2D spectrum consists of four energy branches
pairwise stemmed from the right and left domains. The corresponding
projections are described by energy spectra
$E^{(r)}_{\pm}(k_x=0,k_y)=\beta k_y^2 \pm (\alpha k_y + \Delta_x)$ and
$E^{(l)}_{\pm}(k_x=0,k_y)=\beta k_y^2 \pm (\alpha k_y
- \Delta_x)$~[Fig.~\ref{fig:fig4}.a,b]. The images clearly show the
bound state with the dispersion $\epsilon^{(\parallel)}_B(k_y)$
between the bands $E^{(r)}_{-}$ and $E^{(l)}_{-}$ in the vicinity of
their bottom. The energy band $\epsilon^{(\parallel)}_B(k_y)$ is
symmetric with respect to $k_y=0$ and has a negative effective mass.
When the ratio $\Delta_x/4E_\text{so}$ increases, the maximum value
$\epsilon^{(\parallel)}_B(0)$ decreases. The spin texture of the bound
state $\mathbf{S}(k_y,\epsilon)$ is depicted in
Fig.~\ref{fig:fig4}.c,d. Here, the in-plane component
$S^{(\parallel)}_{xB}(k_y,\epsilon)$ is antisymmetric, hence
$S^{(\parallel)}_{xB}(\epsilon)=0$. At the same time, the out-of-plane
component $S^{(\parallel)}_{zB}(k_y,\epsilon)$ is purely negative and
its value increases with rising $\Delta_x/4E_\text{so}$. As seen in
Fig.~\ref{fig:fig4}.e,f, DOS associated with the bound state,
$\rho^{(\parallel)}_B(\epsilon)$, displays a pronounced peak at
$\epsilon \approx \epsilon^{(\parallel)}_B(0)$.

As mentioned earlier, in the case of the surface divided into two
in-plane domains, the 2D spectrum exhibits the window in the shape of
an improper rhombus around the origin $(k_y=0,E=0)$ restricted by the
curves $E=E^{(r)}_{\pm}(k_x=0,k_y)$ and $E=E^{(l)}_{\pm}(k_x=0,k_y)$
[see Fig.~\ref{fig:fig4}.a,b]. The resonant state is found to occur
inside the window. The corresponding weakly dispersing band
$\epsilon^{(\parallel)}_R(k_y)$ bridges between the nodes with momenta
$(k_x=0,k_y=k^{(\parallel}))$ and $(k_x=0,k_y=-k^{(\parallel)})$. At
the small ratio $\Delta_x/4E_\text{so} \ll 1$, in
Fig.~\ref{fig:fig4} one can observe the emergence of the almost flat
band $\epsilon^{(\parallel)}_R(k_y)$ with a large positive effective
mass and small broadening, which goes over the momentum range in the
1D Brillouin zone, $|k_y| < k^{(\parallel)} \ll \pi/a$. Such features
of the resonant state spectrum are consistent with the analytical
estimations in Eq.~(\ref{eq:5}). With increasing size $\Delta_x$,
the resonant band is up-shifting, its width,
$\epsilon^{(\parallel)}_R(k^{({\parallel})})
- \epsilon^{(\parallel)}_R(0)$, and existence interval,
$2k^{(\parallel)}$, increase as well as the broadening
$\Gamma^{(\parallel)}_R$. When the exchange splitting becomes
comparable to or bigger than the SOC splitting $E_\mathrm{so}$, the
resonant state image blurs and then disappears [see
Fig.~\ref{fig:fig4}.b,d and Fig.~S2]. Strikingly,
Fig.~\ref{fig:fig4}.c,d shows that the resonant state acquires
predominantly spin polarization along the $\mathbf{e}_z$ axis axis
rather than in the plane, especially at small $\Delta_x/4E_\text{so}$.
The polarization sign depends on whether the in-plane DW has a
head-to-head configuration with $S^{(\parallel)}_{zR}(k_y,\epsilon)>0$
or a tail-to-tail one with $S^{(\parallel)}_{zR}(k_y,\epsilon)<0$.
The numerical simulation catches one more resonant state with energy
dispersion $\epsilon^{(\text{III})}_R(k_y)$ located near the edges of
the 2D branches $E^{(r)}_{\pm}(k_x=0,k_y)$ and
$E^{(l)}_{\pm}(k_x=0,k_y)$ at $k_y \approx \pm k_\text{so}$, which can
be seen at $\Delta_x/4E_\text{so}=0.1$. Albeit, at
$\Delta_x/4E_\text{so}=0.4$ the state $\epsilon^{(\text{III})}_R(k_y)$
merges with the state $\epsilon^{(\parallel)}_R(k_y)$.

The above trends are corroborated by the behavior of the total DOS and
the spin density. In the weak exchange field regime
[Fig.~\ref{fig:fig4}.e], we observe three sharp peaks related to the
1D bound states and the two resonant ones induced by DW. With
increasing the size $\Delta_x$ [Fig.~\ref{fig:fig4}.f], the bound
state peak is well maintained, while the resonant states become
smeared. Figure~\ref{fig:fig5} illustrates the spatial profile of the
resonant state $\epsilon^{(\parallel)}_R(k_y)$. Its probability
density is composed of three components: one exponentially decays with
distance from DW, $\sim\exp(\pm k^{(\parallel)}|x|)$, the other
rapidly oscillates, $\sim\cos(4k_\text{so}x+2\phi)$, the third is the
interference of the two formers,
$\exp(-k^{(\parallel)}|x|)\cos(2k_\text{so}x+\phi)$, where
$k^{(\parallel)}= \Delta_x/\alpha < k_\text{so}$, $\phi$ is a phase
shift. At small $\Delta_x/4E_\text{so}$, , the probability density is
strongly concentrated around DW on the scale $|x| \lesssim
[2k^{(\parallel)}]^{-1}$ , where the first component prevails. The
increase of $\Delta_x/4E_\text{so}$ causes a drop in the amplitude and
localization length of the first component [see Fig.~\ref{fig:fig5}].
Thus, if the exchange splitting reaches comparably large values, the
resonant state disappears in both momentum and real space.

\subsection{Noncollinear DW structure}

We now discuss the surface supporting two domains with different
magnetization direction confined in the $(x,z)$ plane, which are
separated by sharp DW at $x=0$. It is convenient to express the
magnetization vectors in spherical coordinates as $\mathbf{M}^{(r,l)}
= M_0 (\sin \vartheta^{(r,l)},0,\cos\vartheta^{(r,l)})$. Then the
sought states generated by the noncollinear DW are specified by a pair
of the angles $\vartheta^{(r)}$ and $\vartheta^{(l)}$ between the
vectors $\mathbf{M}^{(r)}$ and $\mathbf{M}^{(l)}$ and the
$\mathbf{e}_z$ axis, respectively. In the weak exchange field regime,
$\Delta_0/4E_\text{so}=0.2$ ($\Delta_0=JM_0$), Fig.~6 depicts the
redistribution of spectral weight which demonstrates how the prominent
features of the energy structure of the 2D and 1D surface states are
modified under reconfiguration of the DW texture. Evidently, the
noncollinear DW ['($\angle$)'] hosts always the bound state, the
dispersion of which, $\epsilon^{(\angle)}_B(k_y)$, closely follows
modifications in the dispersion of the 2D band near its bottom. In the
case when the out-of-plane components of the domain magnetization
$\mathbf{M}^{(r)}$ and $\mathbf{M}^{(l)}$ have opposite signs, i.e.,
$\cos \vartheta^{(r)} \cdot \cos \vartheta^{(l)} < 0$, as is plotted
in Fig.~\ref{fig:fig6}a,b, the in-gap topologically protected
resonant state with the quasi-linear spectrum
$\epsilon^{(\angle)}_R(k_y)$ is realized. The term ‘topologically
protected resonant state’ is used here in the sense that has been
described in Subsection~\ref{sub:oop_dw}. Further, passing through the
phase with zero exchange splitting in the left domain,
$\vartheta^{(l)}=-\pi/2$ [Fig.~\ref{fig:fig6}.c], the 1D state
$\epsilon^{(\angle)}_R(k_y)$ is transformed into the trivial state,
$\cos\vartheta^{(r)}\cos\vartheta^{(l)}>0$, which is illustrated in
Fig.~\ref{fig:fig6}.d,e. Note, the spectral branch of the trivial
state does not span the gap but clearly originates from the DW. After
the exchange gap in the left domain reverses its sign again,
$\vartheta^{(l)}=\pi/2$ [Fig.~\ref{fig:fig6}.g], the state
$\epsilon^{(\angle)}_R(k_y)$ gets back to the topological character
[Fig.~\ref{fig:fig6}.h]. It should be noted that the weakly damped
fermion excitation at $k_y=0$ propagates along the noncollinear DW at
the velocity $\alpha^{(\angle)}
= \alpha \sin(\frac{\vartheta^{(r)}+\vartheta^{(l)}}{2})$. Depending
on the angles, the velocity may changes not only in the value but also
in the direction in the range of $|\alpha^{(\angle)}| \le \alpha$.

Figure~\ref{fig:fig6}, together with Fig.~S3, gives a hint at the
possibility to control the electron properties by applying external
magnetic field to the system under consideration. For example, in the
case of the Rashba magnetic semiconductor with an easy plane
anisotropy, the magnetization in the domains is tuned from the
in-plane to out-of-plane direction under increasing external field,
$\mathbf{h}=\mathbf{e}_z h$, perpendicular to the surface. At the same
time, the almost flat band $\epsilon^{\parallel}_R(k_y)$ and its
accompanying sharp peak in DOS are shifted by the value
$\Delta_0\cos\vartheta^{(r,l)}$ with respect to the initial position
near zero energy (at $\mathbf{h}=0$) towards higher or lower energy
depending on the direction of the field $\mathbf{h}$. The magnetic
configuration and the corresponding band structure at a value $|h|$ ,
exceeding the anisotropy field but not exceeding the saturation field,
are given in Fig.~\ref{fig:fig6}.d.  Similar
reasoning can be used if external field aligned along the
$\mathbf{e}_x$ direction is applied to the material with the easy axis
anisotropy. In this case, the typical magnetic configuration and
related electron structure are exemplified in Fig.~\ref{fig:fig6}.h.

\section{Discussion}

The surface electron properties of the polar semiconductor BiTeI
possessing the large Rashba splitting $E_\text{so}\approx 100$~meV and
doped with magnetic atoms were thoroughly studied in the
works~\cite{Klimovskikh2017,Shikin_2017+bitei+v,Shikin2021}. The
doping of BiTeI with vanadium had been found to leads to the surface
out-of-plane magnetic ordering and opening of the anomalously large
exchange gap which, according to the ARPES
measurements~\cite{Klimovskikh2017, Shikin_2017+bitei+v, Shikin2021},
reaches $2\Delta_z=125$~meV at the Kramers point at the
V-concentration of 2\% and temperature 20~K. Interplay of the
extraordinary large Rashba splitting and the sizable exchange gap
would be ideally suited to satisfy the requirements for the
realization and control of spin-polarized conductivity. However, in
the Bi$_{1-\mathrm{x}}$V$_\mathrm{x}$TeI material the situation has its own
specifics. Initially, when no magnetic field is applied, the observed
total out-of-plane spin polarization of the samples is turned out to
be very weak which does not match such a large gap. The authors of
Ref.~\citenum{Shikin_2017+bitei+v} rationalized this seeming
contradiction by assumption of the spontaneous formation of the
surface ferromagnetic domains of almost equal area which are
oppositely oriented below the Curie temperature $T_c=130$~K.  As was
shown above, the out-of-plane DW induces the partially polarized bound
state beyond the band continuum. What is even more interesting is that
the chiral resonant state with Dirac-like dispersion exists inside the
exchange gap. The quasi 1D unidirectional channel localized along DW
can manifest itself, provided that the Rashba splitting energy exceeds
the exchange one. This channel carries a weakly decaying fermion
excitation, surviving for relatively long time $\tau^{(\perp)} \sim
1/\Gamma_R^{(\perp)}$. During this time, in a ballistic transport
regime with the Fermi energy level being within the gap, $|\mu|
< \Delta_z$, the excitation passes the distance
$l_R^{\perp)}=\alpha\tau^{(\perp)}_R$. Authors of
Ref.~\citenum{Shikin_2017+bitei+v}, studied the dependence of the
surface exchange gap on temperature $T$ and concentration $\mathrm{x}$ of magnetic
impurities via the ARPES measurements in V- and Mn-doped BiTeI
samples. For example, at $\mathrm{x} \approx 2$\% and $T=100$~K, they estimated
gap as $2\Delta_z \approx 80$~meV. This means that
$\Delta_z/4E_\text{so} \approx 0.1$, hence, according to the
dispersion relation of Eq.~(\ref{eq:4}), the broadening is about
two orders of magnitude smaller than the gap,
$\Gamma^{(\perp)}_R \approx 2\Delta_z \cdot 10^{-2}$. Taking into
account that BiTeI has Rashba parameter
$\alpha_R=3.8$~eV$\cdot$\AA~\cite{Ishizaka2011}, the characteristic
length $l^{(\perp)}_R$ can be roughly evaluated as large as 50~$\mu$m.
The latter even surpasses the inelastic scattering length observed in
the TI thin film Cr-doped (Bi,Sb)$_2$Te$_3$ in the QAH regime at low
temperatures~\cite{Deng2022}. Of course, the detailed analysis should
include other dissipation mechanisms connected with the bulk bands and
mid-gap impurity states as well as random magnetic multi-domain
structure, which may reduce the length $l^{(\perp)}_R$.

For semiconductors with the strong Rashba effect, experimental
information about the surface properties modification triggered by DW
still lacks so far, which would motivate experimentalists to perform
relevant investigations. The developed theory may suggest adequate
experimental approaches to test the signatures of the peculiar surface
states in these materials. The combination of scanning tunneling
microscopy and spectroscopy (STM/STS) and magnetic force microscopy
(MFM) provides powerful tool for this purpose. In the virgin phase a
sample possesses a multi-domain structure~\cite{PhysRevX.8.021067,
Shikin_2017+bitei+v}. MFM can select a single DW on the sample surface
and identify its vector and spatial configuration, while STM/STS can
map the electron density surrounding the DW through differential
conductance measurements. We consider the surface of BiTeI doped with
transition metal atoms as the object of such a study. When the Fermi
level lies lower (higher) than or near to the 2D conduction band
bottom (valence band top), STM/STS provides the signature of the bound
state $\epsilon^{(\perp,\parallel)}_B(k_y)$, which is shifted towards
the bigger binding energy with increasing the exchange gap size
$2\Delta_z(T)$, i.e., with decreasing temperature $T$. By tuning the
Fermi level to the exchange gap, the tunneling conductivity can be
locally enhanced in a narrow stripe stretching along DW if, as argued
above, the evanescent component of the envelope function of the
resonant state $\epsilon^{(\perp)}_R(k_y)$ dominates over the extended
one. Moving away from the DW at the distance longer than about $\sim
[k^{(\perp)}]^{-1}$, the STM/STS signal significantly decays. This
situation could be observed in the Bi$_{1-\mathrm{x}}$(V,Mn)$_\mathrm{x}$TeI samples
under the optimal doping $\mathrm{x}\approx$~2--3\% over the temperature region
where the magnetic order exists. Indeed, in accordance with the
dependence $2\Delta_z(T)$ deduced in Ref.~\citenum{Shikin2021}, the
condition of the relatively weak exchange energy is satisfied at
$0<T<T_c$. Thereat, it is hard to detect a trace of the resonant state
in close proximity to $T \approx T_c$ where the Fermi-Dirac distribution
broadening surpasses the gap. Thus, there are reasons to regard the
surface of diluted magnetic semiconductor on the base of BiTeI as a
possible platform for realizing the quasi-robust chiral conducting
channels at rather high temperatures, above $\sim$100~K.

In the system with out-of-plane anisotropy, both the 1D resonant
states hosted at DWs and the lower subband of the 2D Rashba state
simultaneously contribute to the surface electric transport when
$|\mu|<\Delta_z$. In this energy range, the presence of a
magnetization has weak effect on the 2D subband $E_{-}(\mathbf{k})$.
The existence and characteristics of the resonant state are sensitive
to the surface magnetic moments arrangement. In turn, the latter can
be driven by an external magnetic field. Figures 6 and S3 demonstrates
how the resonant state spectrum depends on the orientation of the
domain magnetizations with respect to each other and to the surface
normal. Furthermore, in a field-sweep experiments, during a
magnetization reversal process under the field
$\mathbf{h}=\mathbf{e}_z h$ the concentration of DWs increases,
reaching a maximum at the coercive field $h_c$. Correspondingly, the
concentration of the in-gap resonant states crossing the Fermi level
increases. Since the magnetic domain scale is assumed to be much
larger than the resonant state localization length $\sim
[k^{(\perp)}]^{-1}$, the sample surface can be considered as a network
of the 1D percolating channels. Therefore, an enhancement of a
longitudinal conductivity $\sigma_{xx}(h)$ can be detected in the
vicinity of the field $H_c$, provided the resonant state contribution
dominates over the 2D subband one. Similar behavior of
$\sigma_{xx}(h)$ has been disclosed and substantiated in TI
Bi$_2$(Te,Se)$_3$ doped with Mn~\cite{Checkelsky2012}.  In contrast,
typically ferromagnets show a decrement of the longitudinal
conductivity near the coercive field associated with the increased
electron scattering at DWs~\cite{PhysRevLett.79.5110}. Thereby, we
predict that the Rashba magnetic semiconductor with an easy-axis
anisotropy can manifest an peculiarity in magnetoresistance.

As for the resonant state with the flattened highly spin-polarized
band $\epsilon^{(\perp)}_R(k_y)$ caused by the in-plane DW, it gives
rise to a sharp DOS peak in the vicinity of $E=0$, which can be fairly
easily identified by STM/STS. Albeit, whether such a singularity can
enhance the inter-electron interactions to provoke instabilities at
the DW (such as charge density wave or superconductivity) is an open
question. To verify it one needs to fabricate the Rashba material
possessing an easy plane magnetic anisotropy.

\section{Conclusion}

In summary, we present a theoretical investigation of electron states
induced by an inhomogeneous magnetization at the surface of a
semiconductor with the strong Rashba splitting. We predict that a
single magnetic DW hosts not only the bound state but also, what is
especially noteworthy, the 1D chiral resonant state. The properties of
the resonant state depend profoundly on the characteristic parameters
of the model and the magnetization orientation in adjacent
domains. With decreasing ratio of the exchange energy to the Rashba
splitting energy, the resonance broadening gets narrower and the state
becomes more localized at DW. The dispersion relation ranges from a
linear spectrum that tends to cross the local exchange gap to almost
flat-band spanning the Kramers degeneration nodes when the domain
magnetization varies from out-of-plane to in-plane. We substantiate
the physical reason why the 1D chiral resonant state is guaranteed to
appear at the antiphase DW. Our finding expands the realm of systems
in which the convergence of magnetic order and SOC leads to the
creation of peculiar electron states. The estimations demonstrate that
the weakly damped resonant states can be materialized at the pristine
cleaved surface of BiTeI doped with transition metal atoms and their
signatures can be sought out in the magnetotransport and STM/STS
studies. The unique surface electron states, whose possibility we
predict in diluted magnetic semiconductors with strong Rashba
splitting, being manipulated by both electrostatic and magnetic means
open new opportunities for use in spintronic devices.

\section{Acknowledgements}

I.P.R acknowledges financial support from the Ministry of Education
and Science of the Russian Federation within State Task
No. FSWM-2020-0033 (in the part of tight-binding calculations). E.V.C
acknowledges Saint-Petersburg State University for a research project
95442847.

\bibliography{bibfile}

%\section{THE MAIN RESULTS}
%\subsection{Single antiphase domain wall}
\end{document}